\documentclass[a4,12pt]{article}

\usepackage{graphicx}

\usepackage{authblk}

\usepackage{algorithm}

\usepackage{algorithmic}

\usepackage[figuresright]{rotating} 

\usepackage{epsfig,times,latexsym,enumerate,lscape,flafter,amsmath,amssymb,graphicx}
\usepackage{blindtext}
\usepackage{graphicx}
\usepackage{adjustbox}
\usepackage{cite}
\usepackage{amsmath}
\usepackage{amsfonts}
\usepackage{lipsum}
\usepackage[bookmarks=false]{hyperref}

\begin{document}

\title{An investigation of a deep learning based malware detection system}

\author{Mohit Sewak, Sanjay K. Sahay\thanks{E-mail: ssahay@bits-goa.ac.in} \hspace{0.15cm} and Hemant Rathore\\ 
Department of Computer Science and Information System, Birla Institute
of Technology and Science, K. K. Birla Goa Campus, NH-17B, By Pass Road, Zuarinagar-403726, Goa, India
}
\date{}

\maketitle

\begin{abstract}

We investigate a Deep Learning based system for malware detection. In the investigation, we experiment with different combination of Deep Learning architectures including Auto-Encoders, and Deep Neural Networks with varying layers over Malicia malware dataset on which earlier studies have obtained an accuracy of (98\%) with an acceptable False Positive Rates (1.07\%). But these results were done using extensive man-made custom domain features and investing corresponding feature engineering and design efforts.  In our proposed approach, besides improving the previous best results (99.21\% accuracy and an False Positive Rate of 0.19\%) indicates that Deep Learning based systems could deliver an effective defense against malware. Since it is good in automatically extracting higher conceptual features from the data, Deep Learning based systems could provide an effective, general and scalable mechanism for detection of existing and unknown malware.\\
\noindent {\bf Keywords:} Malware, Deep Learning, Machine Learning, Auto Encoders, Deep Neural Networks, Malicia, Cyber Security.
\end{abstract}

\maketitle

\section{Introduction}
Various attempts have been made in the past to detect and classify malware from benign programs. These methods range from the early-day signature-based detection to the more modern Machine Learning and Deep Learning based detection.

Besides the detection techniques, the features used and data processing also plays a major role in such studies. Some researchers suggest that opcode frequency-based detection is superior to signature-based detection \cite{santos2013}. The study by Santos et al. 2013 claims to have obtained an accuracy of 95.9\% using opcode frequency as features.

Later on, researchers also applied file-size based segmentation on the opcode frequency and improved the average accuracy over different Machine Learning models slightly above 96\% (96.28\%), and claimed that Random Forest provided the best individual accuracy close to 98\% (97.95\%) with a False Positive Rate (FPR)) of 1.07\% \cite{ashu2016} using the malware samples from the Malicia project data \cite{malicia}.

We did not use any custom feature engineering techniques like the segmentation method as used in the above work, instead, we use the opcode frequency directly as used in the study by Santos et al. \cite{santos2013}. We used different architectures of Auto-Encoders (AE) for feature extraction (along with scaling the data) and achieved a much better accuracy of 99.21\% with even a better False Positive Rate (FPR) of 0.19\% using just a 4-layer Deep Neural Network (DNN) for classification with a 3-layer AE for feature extraction.

Our work indicates that Deep Learning based architectures such as Auto-Encoders (for Feature Extraction) and Deep Neural Networks (for malware classification) may provide a very effective system for defense against many types of malware, that too without devising any custom features preparation techniques, yet provide much better accuracy that too at a much lower FPR.

The remainder of this paper is organized as follows. Section 2 introduces the different approaches used for malware detection in the past. Section 3 explains the Malicia project and the malware and benign files used, and section 4 covers the approach for converting these executable into structured data for deep learning application. Section 5, and 6 covers the deep learning approaches for feature extraction (AE) and modeling (DNN). The results are stated and analyzed in Section 7, and some conclusions and ideas for future work are given in Section 8.

\section{Related work on Malware Detection}
There are two types of approaches to identify a MALWARE (or a MAL-icious soft-WARE) from a benign software, namely the \textit{static analysis} and the \textit{dynamic analysis}. 

In the static approach, signatures or other important static features \cite{kolter2004learning} \cite{schultz2001data} are extracted from the software file or its associated meta-data without running or executing it and then the detection algorithm is applied on these extracted signatures or features. Some examples of such features are Portable Executable, strings and byte sequences \cite{schultz2001data}, and function length frequency. Whereas in the dynamic analysis, the software is run or executed, and then important behavioral aspects and features are captured, and the detection algorithm is run on such behavioral features and observations.

There also exists approaches where multiple static, dynamic or hybrid approaches in which features from both static and dynamic analysis could be combined, and there also exists methods like binary/ opcode analysis which could fall in either or both type of analysis depending upon how involved the feature extraction method is. Usage of opcode based features started gaining popularity. There is some work in which directly opcode frequency distribution has been used \cite{ashu2016} \cite{bilar2007opcodes} \cite{ashu2018} \cite{ye2017} \cite{ahmadi2016}, whereas in others the opcode sequences and permutations have been found useful \cite{karim2005}.

Direct signature-based and simple static methods have a lot of shortcomings, especially when used for detecting the second generation Polymorphic, and Metamorphic malware, and hence most active research has migrated to more sophisticated techniques like Machine Learning and Deep Learning.

In the Machine Learning and related approaches, typically a supervised learning or classification algorithm is applied on the collected features for detection (binary model for differentiating between malware and benign software) or classification (multinomial model for distinguishing across the family of malware). Some examples of such supervised models are instance-based learning algorithms \cite{kolter2004learning}, TF-IDF \cite{santos2013} \cite{kolter2004learning}, Naive Bayes \cite{kolter2004learning}, Support Vector Machine \cite{santos2013} \cite{kolter2004learning}, Decision tree \cite{kolter2004learning}, ID3 \cite{henchiri2006feature}, J48 \cite{henchiri2006feature} \cite{ashu2016}, Hidden Markov Models \cite{attaluri2008}, Random Forest \cite{ashu2016} and AdaBoost \cite{tabish2009malware}.

Optionally, instead of applying the supervised learning model directly on the collected features, some sort of feature selection \cite{lin2015}, feature engineering or feature/dimension reduction \cite{dahl2013} algorithms could be additionally used before supervised learning. A popular way for feature engineering is to use an unsupervised learning algorithm for making better representative features for the supervised learning stage. Some example of feature selection methods used are hierarchical \cite{henchiri2006feature}, unary variable removal \cite{siddiqui2008detecting}, Goodness evaluator \cite{mehdi2009imad}, and Weighted Term Frequency \cite{santos2013}.

Some drawbacks of conventional supervised learning methods are that a lot effort and domain expertise is required in creating and identifying important features \cite{Xin-2016}, high false positive rates from such techniques, and their assumptions are too simple (naive) to provide higher conceptual features \cite{Pascanu-2015}. Due to these shortcomings, Deep Learning based techniques are gaining a lot of traction in malware detection.

A lot of recent efforts has been focused on and success achieved with different Deep Learning techniques like Deep Belief Networks \cite{DeepSign-2015}, Deep Neural Networks \cite{Saxe-2015}, Recurrent Neural Networks (RNN) and its variants like Long Short-Term Memory (LSTM) and Gated Recurrent Unit (GRU) \cite{Pascanu-2015}  and combination of recurrent and convolutional neural networks \cite{Kolosnjaji-2016} for supervised learning and classification. Similarly, for feature engineering and lower dimensionality feature extraction Deep Belief Networks (DBN) and (Stacked) Auto-Encoders (SAE/ AE), and RNN based Auto-Encoders (RNN-AE) \cite{Xin-2016} have been used.

\section{Malware Data Used}
Malware data was collected from the Malicia project. The project contains almost 11,368 malicious files, from which assemble codes were extracted using the Unix's \textit{objdump} utility which succeeded in generating the assemble (*.asm) files for 11,084 files.

Another sample of 2819 benign executable files was collected from different windows systems and `Cygwin' utility, and each file in the sample was verified to be non-malicious using the \textit{virustotal.com} file scanner services. This dataset with almost 11,000 malicious and approx. 2,800 benign samples, leads to class imbalance problem, which was resolved using the Adaptive Synthetic (ADASYN) oversampling technique

\section{Structured Data Generation}
From the generated Assemble Code file for each malicious and benign executable, their unique opcodes were extracted and a such extracted unique opcodes made. The set was converted into an index, where each unique opcode was assigned an integer. A total of over 1600 unique opcodes were indexed this way (we will call this list as the `master opcode list').

Next using the index of the master opcode list, the frequency of occurrence of each of the available opcode was measured in each assemble file and recorded against them in a single row in a file called opcode frequency (this is done separately for benign and malicious files, and then merged into a single file). Additionally, a label (0 or 1) corresponding to whether the data in a given row is from a benign (0) or a malicious (1) executable.

Further, these frequencies were scaled between $[0,1]$. We use the scaled frequencies of occurrence of each code in each file (as X features) with the respective file labels (as Y labels) as our dataset, which was further split into training (2/3rd) and test (remaining 1/3rd) data splits randomly. This training and test sets were used as described in the following sections on data processing and modeling.

\section{Feature Extraction/ Dimensionality Reduction} 
Two different variants of (Stacked) Auto-Encoders were used for the purpose of feature extraction, or dimensionality reduction. These were:
\begin{itemize}
\item
A single layer (1-encoder, 1-decoder for training) Auto-Encoder, here referred as 1-layer AE (or Non-Deep Auto-Encoder as named here).
\item
A 3-layer (Stacked) Auto-Encoder (3-encoders, followed by 3-decoders for training), here referred as 3-layer AE (or Deep Auto-Encoder as named here).
\end{itemize}
Auto-Encoders are unsupervised learning algorithms and do not require the actual labels for training, hence only the X features (opcode frequencies).

All Auto-Encoder layers use Exponential Linear Units (ELU) as activation except for the last layer which uses linear activation as we intend to predict a continuous variable to reconstitute the input. The bottleneck layer in both auto encoders contains 32 nodes. The 3-layer AE additionally use 128, and 64 node hidden layers in encoder and decoder (Architecture/Node-count: Input-128-64-32-64-128-Output), whereas, in 1-layer AE, the encoder/ decoder directly connects the bottleneck layer (32) to input/ output layer (Architecture/Node-count: Input-32-Output).

Mean square error, which is the mean of the squared error over samples in a batch, has been used as loss function for training the Auto-Encoders as the scaled features represent continuous variables. Both AE here use ADAM optimizer and were trained for 120 epochs with a batch size of 64 samples in each batch.

The corresponding figure (Figure 1) shows the training and validation (test) losses across epochs during the training of one of the Auto-Encoders.


\begin{center}
\begin{figure}
\includegraphics[width=3.15in,height=6in,clip,keepaspectratio]{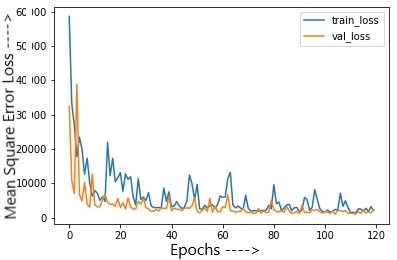}
\caption{Sample plot showing training and validation loss (y-axis) across each epoch (x-axis) during training of an AE.}
\end{figure}
\end{center}

\section{Modeling} 
Three Deep Neural Network models were used with features extracted from both of the above Auto-Encoders. These models are:
\begin{itemize}
\item A 2-hidden layer Deep Neural Network, here named as 2L-DNN.
\item A 4-hidden layer Deep Neural Network, here named as 4L-DNN.
\item A 7-hidden layer Deep Neural Network, here named as 7L-DNN.
\end{itemize}
All deep neural networks layers use ELU as activation except for the last layer which uses softmax activation. 

All deep neural networks use ADAM optimizer, binary cross entropy loss function, and 0.1 dropout \cite{nitesh2014} in each of the hidden layer. Since we require a binary classification, therefore the output layer contains only 1 node with Sigmoid activation. All the other layers use ELU activation. All the DNNs were trained for 120 epochs, with a batch size of 64 samples.

The 7-layer DNN (7L-DNN) contains $2^{11-i}$ nodes ($1024, 512, ..., 16$ nodes) in the $i^{th}$ hidden layer (as counted from the input layer), and the 4-layer DNN (4L-DNN) contains $2^{12-(2\times i)}$ nodes in the $i^{th}$ hidden layer ($1024, 256, 64, 16$ nodes).
The 2-layer DNN (2L-DNN) contains 1024, and 32 nodes in the respective 2 hidden layers.

Below is a sample plot (Figure 2) showing a loss as a function of epochs across the training iterations. After each training, the model is tested on the test set, and corresponding validation losses have been computed at each epoch. The asymptotic nature of these plots indicates convergence which suggests that the training is satisfactory with respect to the used hyper-parameters. 

One observation with respect to this plot is that at occasions the training loss is more than the validation loss, which seems counter-intuitive at first, but there are two explanations for this. First is that we use dropout (regularization) during training [23], which is removed during validation, this though avoids over-fitting but increases the training loss as many neurons are 'off' during training randomly. The second reason is that the training loss is average across the samples in a training batch, whereas the validation loss is computed after the weights are optimized in each epoch for that epoch, this gives an impression of lower validation loss as compared to training loss for a given epoch.

\begin{figure}
    \centering
    \includegraphics[width=3.15in,height=6in,clip,keepaspectratio]{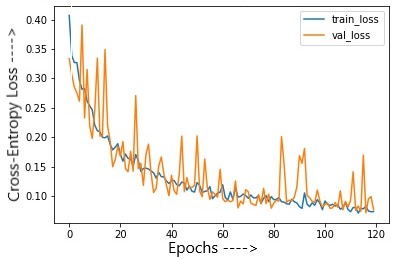}
    \caption{Sample plot showing training and validation loss (y-axis) across each  epoch (x-axis)  during training of a DNN.}
\end{figure}

\section{Results}
The table below (Table 1) shows the True Positive, False Positive, True Negative and False Negative instance counts with different AE and DNN combination.

\noindent Where,
\begin{itemize}
    \item TP = true positives; malware identified as malware.
    \item FP = false positives; benign identified as malware.
    \item TN = true negatives; benign identified as benign.
    \item FN = false negatives; malware identified as benign.
\end{itemize}

\begin{table}[ht]
\centering
\small
\begin{tabular}{|l|l|c|c|c|c|}
\hline
\hline
AE & DNN & TP & FP & TN & FN \\
\hline
\hline
1L-AE & 2L-DNN & 3481 & 23 & 3612 & 200 \\
1L-AE & 4L-DNN & 3451 & 44 & 3591 & 230 \\
1L-AE & 7L-DNN & 3618 & 11 & 3624 & 63 \\
\hline
3L-AE & 2L-DNN & 3630 & 157 & 3478 & 51 \\
3L-AE & 4L-DNN & 3630 & 7 & 3628 & 51 \\
3L-AE & 7L-DNN & 3238 & 25 & 3610 & 443 \\
\hline
\hline
\end{tabular}
\caption{Performance results (instance counts) with different combinations of AE and DNN.} 
\end{table} 

The Table 2 shows the performance evaluation (on test set) for the different combination of AE and DNN used. It gives the False Positive Rate (FPR), True Positive Rate (TPR), True Negative Rate (TNR), Positive Predictive Value (PPV) and the Accuracy (Acc.).
where, \hfill
\begin{itemize}
\item $FPR = FP/(FP+TN)$ 
\item $TPR = TP/(TP+FN)$; also denoted as \textbf{Sensitivity}, \textbf{Recall} or \textbf{Hit Rate}.
\item $TNR = TN/(TN+FP)$; also denoted as \textbf{Specificity}.
\item $PPV = TP/(TP+FP)$; also denoted as \textbf{Precision}.
\item $Accuracy = \frac{TP+TN}{TP+TN+FP+FN}\times100$\%
\end{itemize}
\begin{table}[ht]
\centering
\small
\begin{tabular}{|l|l|c|c|c|c|c|}
\hline
\hline
AE & DNN & FPR & TPR & TNR & PPV & Acc. \\
\hline
\hline
1L-AE & 2L-DNN & 0.0063 & 0.9457 & 0.9937 & 0.9934 & 96.95 \\
1L-AE & 4L-DNN & 0.0121 & 0.9375 & 0.9879 & 0.9874 & 96.25 \\
1L-AE & 7L-DNN & 0.0030 & 0.9829 & 0.997 & 0.997 & 98.99 \\
\hline
3L-AE & 2L-DNN & 0.0432 & 0.9861 & 0.9568 & 0.9585 & 97.16 \\
3L-AE & 4L-DNN & 0.0019 & 0.9861 & 0.9981 & 0.9981 & 99.21 \\
3L-AE & 7L-DNN & 0.0069 & 0.8797 & 0.9931 & 0.9923 & 93.60 \\
\hline
\hline
\end{tabular}
\caption{Performance results (rates) with different combinations of AE and DNN.} 
\end{table} 
Summarizing the findings from above tables, we find from the below Accuracy (Figure 3) and FPR (Figure 4) plots that 3-layer AE with 4-layer DNN gives the best combination of accuracy (99.21\%) and FPR (0.19\%) simultaneously. On the other hand, 3-layer AE with 7-layer DNN offered the lowest accuracy, though at an FPR of 0.69\% it is not only comparable to but also much better than that obtained in previous studies.
\begin{figure}
    \centering
    \includegraphics[width=3.15in,height=6in,clip,keepaspectratio]{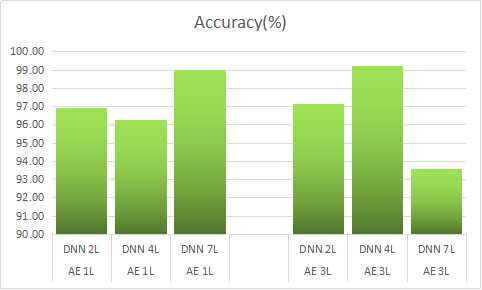}
    \caption{Accuracy for different combinations of AEs and DNNs.}
\end{figure}
\begin{figure}
    \centering
    \includegraphics[width=3.15in,height=6in,clip,keepaspectratio]{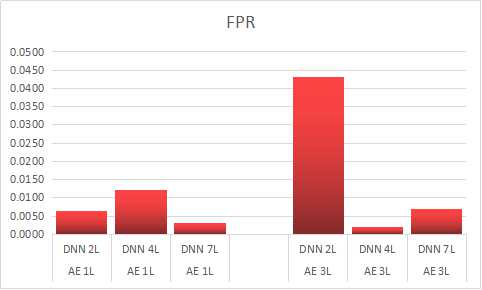}
    \caption{False Positive Rates for different combinations of AEs and DNNs.}
\end{figure}
One interesting finding from these plots could also be that combining a deeper 3-layer AE with shallower DNNs (2-layer and 4-layer) provides much better results than when combined with deeper DNN (7-layer), which is exactly the opposite trend when the shallower 1-layer AE is combined with deeper 7-layer DNN. In-fact the 7-layer DNN gives the worst performance with 3-layer AE and the best performance with 1-layer AE. One explanation for this finding could be that a combination of complex-features (deeper the networks, more complex could be the function fit) function and complex-classifier function is leading to overfitting. This trend manifests itself, to also enable the simpler 1-layer AE to give the second best performance both from accuracy perspective (\~99\%) and FPR perspective (0.30\&) when combined with the deepest DNN (7-layer).

\section{Conclusion}
To the best of our understanding, this is the first study that could produce such results where we improved the previous best accuracy by 1.26\% (99.21\% as compared to 97.95\%) while simultaneously also improving FPR by 5.6 times (0.19\% as compared to 1.07\%) on the Malicia dataset. The results become more important because we have just followed the suggested best practices for Deep Learning, and besides that, we did not create or implemented any expert-designed custom features as was done in some of the previous studies that improved prediction accuracy and FPRs on Malicia data earlier.

Since we did not create any custom domain features and had used only the opcode frequency directly (as-is) for feature selection/ extraction and classification, a more direct comparison could be with the best results under similar conditions of using opcode frequency directly. Under these constraints, we took even a larger leap and improved the accuracy by 3.31\% (99.21\% as compared to 95.9\%) while simultaneously improving FPR by 10.5 times (0.19\% as compared to 2.0\%) from the previous best results of studies done (using opcode frequency directly) on Malicia data.

Another contribution of our investigation is comparing different combinations of Auto-Encoder and Deep Neural Network (of varying layers) combinations to see how effective each combination is, and surprisingly the best combinations did not come from either of the extremes (very deep or shallow networks).

This investigation goes a long way to prove that effective implementation of end-to-end deep learning architectures from feature extraction to classification could provide very credible, general and scalable defense against many types of malware, without requiring any special feature engineering requirements. This aspect of deep learning to automatically extract higher conceptual features from data is very useful in making a general malware defense mechanism that could work against many malware families and types without much incremental effort.

Since some of the previous works had already attained very good (over 95\%) accuracy with Malicia dataset, as a proposal for further work, we believe that using a more complex dataset of Meta/Poly-morphic malware could help further differentiate deep learning based Malware defense mechanisms against previous approaches. This is evident from the fact that the data was not complex enough to get the best out of the deepest AE and DNN combination in the study. 
Further, it will also be interesting to see how different other deep learning techniques and architectures like RNN, LSTM, ESN, etc. perform, especially when coupled with the more advanced data processing and feature engineering methods to enable a comprehensive malware defense system.
\bibliographystyle{unsrt}
\bibliography{ares18-PaperID73}

\end{document}